# Optimizing Character Animations using Online Crowdsourcing

Benjamin Kenwright

## Abstract

This paper presents a novel approach for exploring diverse and expressive motions that are physically correct and interactive. The approach combining user participation in with the animation creation process using crowdsourcing to remove the need for data-driven libraries while address aesthetic appeal. A core challenge for character animation solutions that do not use pre-recorded data is they are constrained to specific actions or appear unnatural and out of place (compared to real-life movements). Character movements are very **subjective** to human perception (easy to identify unnatural or strange patterns even within simple actions like walking or climbing). We present an approach that leverages crowdsourcing to reduce these uncanny artifacts within generated character animations. Crowdsourcing animations is an uncommon practice due to the complexities of having multiple people working in parallel on a single animation. A web-based solution for analysis and animation is presented in this paper. It allows users to rate optimizations and view character animation details conveniently on-line. The context of this paper introduces a simple animation system, which is integrated into a web-based solution (JavaScript/HTML5). Since Web-browsers are commonly available on computers, the presented application is easy to use on any platform from any location (easy to upgrade, maintain and share). Our system combines the expressive power of web pages for visualizing content on-the-fly with a fully-fledged interactive (physics-based) animation solution that includes a rich set of libraries.

## 1 Introduction

**Expressive Animations without Key-Frame Data** Creating high quality character animations is an expensive and challenging task. Virtual characters are able to perform a paradigm of both physically correct and expressive motions that embody both emotions and actions. For instance, articulated biped characters are capable of a vast assortment of posses and styles, even the mundane task of balanced upright walking has limitless possibilities, as demonstrated by the The Ministry of Silly Walks [1] which shows an assortments of creative walking styles. If we consider the kinds of animations created by Disney and Pixar (some of the highest standards of animation quality on the planet). On average, a 300-600 man/woman strong studio would over a period of three years to produce a fully featured animated film. The basics of animation are simple. The complexities are in achieving a high quality solutions, i.e, creating natural characters, captivating and compelling animations which appear organic and realistic (alive). The back and forth process of tuning and refining animations can last over 12 months (this does not include the actual animation of the scene and environment) [2–4].

**Movies to Interactive Worlds** This is compounded by character animation solutions that are used in interactive environments, which need to react and adapt to unforeseen circumstances in a natural and physically-correct way. Currently animations are created using pre-recorded data or by artists. This paper explores a novel way for creating high-quality character animations without pre-recorded data and reduced artistic burden. Immersive interactive environments are pushing the boundaries of our imagination, not to mention, the expectation standards. Animated characters are required to be life-like and dynamic and must go beyond pre-recorded puppet-like figures. Characters that adapt and move in a natural way based on the situation conditions (pushed or walking up stairs). Animation creation and edition and is an expensive business (time consuming, computationally and quality).

**Previous approaches** As mentioned earlier, since the majority of high quality character animation techniques are dependent on large data libraries, such as, high quality human motion capture, there has been a substantial amount of research into this area (i.e., adapting and working with these large data-sets) compared to non-standard approaches that try to create life-like animations without any key-frame data (as proposed in this paper). Ultimately this is because data driven methods continue **generates the most realistic solutions** [5]. Of course, data-driven methods are thwart with challenges, such as interactive manipulation [6], re-targeting motion to new characters [7], blending a family of similar motions [8], statistical modeling [9], and incorporating physics-based objectives and constraints [10]. The also includes dealing with multiple characters in the context of character-character interaction and manipulation [11]. These approaches have been merged with physically-based solutions that use optimizations and hybrid concepts (for offline training and machine learning) but again require training data to create natural motions. The central problem behind purely procedural (or algorithmic approaches) that do not depend on data-driven solutions is they are either limited to specific actions and skeleton topology (biped walking or standing) or are prone to aesthetic limitations - the problem of unnatural motions (uncanney valley) [12]. This is most prominent when animations are required to adapt to unforseen circumstances in a natural life-like manner (interactive environments). Previously, there has been a wide gulf between animations produced by commercial organisations and those by independent sources. Through crowdsourcing, we may see a narrowing of this gap with a reduced cost while continuing to push the quality standard constraints.

**Our approach** As far as we are aware, this paper is the first to approach this problem using the proposed solution. **Help identify when an animation movement is uncomfortable or uncanny**. Combining online web-based technologies with crowdsourcing to tune optimized animations to be more natural and life-like (without training data). This is especially important for interactive environments, which require autonomous animation solutions which are able to react and adapt in real-time in an organic manner. We believe our approach works well for crowdsourcing. As crowdsourcing works especially well for certain kinds of tasks, typically, ones that are fast to complete (or evaluate), incur low cognitive load, have low barriers to entry, are objective and verifiable, require little expertise, and can be broken up into independent sub-tasks. As users are ideal for evaluating aesthetic qualities and identifying uncanny artifacts within generated animations (i.e., volunteers are able to watch short animation segments and rate the quality and naturalness quickly and easily without any specialist skills or training).

**Contribution** The key contributions of this paper are: (1) combines automated online character optimization techniques with crowdsourcing to address aesthetic limitations; (2) web-based implementation that is able to run client-side for simulating and testing animation solutions (customizable features and options, such as, the skeletal configuration, optimization algorithm and so on); (3) animation solution does not require or depend pre-recorded animation data (motion capture libraries or pre-created key-frame references).

## 2 Related work

**Character Animation** Inverse kinematics solutions are able to map articulated character posses to accomplish coordinated and controlled motions [13, 14], such as, foot at specific locations and hands-fingers when opening doors or avoiding obstacles. However, inverse kinematic are rarely used for creating full posture motions by themselves as they lack the fluid-like movements inherent in organic creatures - and are instead combined with other approaches, such as, physical models and data-driven methods (e.g., adapting key-frame animations).

Physics-based models are able to simulate accurate real-world character representations (weight, mass, joints and contacts) [15]. These physical representations enable the character models to be responsive and interactive (react automatically to collisions and hits). While physical simulation guarantees realistic interactions in the virtual world, simulated characters can appear unnatural if they perform unusual movements [16].

Data-driven motions (puppets) are a popular choice for controlling character animations as the data can be taken from real-world recordings of humans - containing subtle details which make the motions more natural and life-like. Inverse dynamics [15] enables the motions to be used to calculate the control forces for the physics-based skeleton (akin to the strings controlling the puppet). Of course, purely data-driven solutions have limitations, and are highly dependent on the animation library, any new actions or movements would require new recordings, also the shape and skeleton topology, not to mention, the react of animations to on-the-fly disturbances is undefined and need to be integrated in somehow. For example, mechanistic approaches can be used to address specific actions, like walking and balancing [17, 18] so the character reacts to different terrain and forces naturally.

Optimization based-methods have gained a lot of attention recently as a promising approach for creating **new** character animations. These methods have been accelerated and combined with hardware solutions to address long computational times (e.g., Kenwright [19] accelerated the process using parallel computing). Hybrid techniques have attempted to combine inverse kinematics and machine learning [20] to address ambiguities and allow artists more control. However, ultimately these approaches are either highly dependent on data to create solutions which are natural and organic or a considerable amount of artistic input. As trained artists have a skilled eye for identifying issues and providing input/tweaks to the trained solution, usually through animation data/input/constraints.

**Optimizing Neural Networks** Optimization algorithms for training neural network topologies originated from the field of evolutionary computing in the 1990s [21–23], although its origins can be traced back to Alan Turing's Unorganized Machines [24]. Our method adds crowdsourcing to evaluate solution, with optimization searches performed locally on the client machine (not dependent on the internet connection for tuning the network weights). Machine learning has been a popular approach for developing adaptable animation solutions using prerecorded training data [25, 26]. However, these approaches are highly dependent on large animation databases for training the solution.

**Crowdsourcing** Crowdsourcing is a problem-solving approach using the collective intelligence of networked communities [27]. Crowdsourcing has been successful in design [28], education and even cybersecurity [29]. However, despite crowdsourcing being a growing practice in many companies in different sectors, it remains little understood or utilized in the creative sector (e.g., graphics and animation) [30]. We use crowdsourcing in this paper as a means to add humans-into-the-loop to find solutions to a difficult and complex problem. A problem that is very easy for humans to solve but difficult for machines (that is, to identify unnatural motions). As people in general are very good at identifying small anomalies or glitches in animations that would be near impossible using purely algorithmic means, as explained by the uncanny valley phenomenon [12]. The uncanny valley describe the observation that as motions appear more humanlike, they become more appealing, but only up to a certain point. Getting past this point is often very difficult, as even the smallest error or mismatch with the coordinated full body motion can result in an unnatural perception.

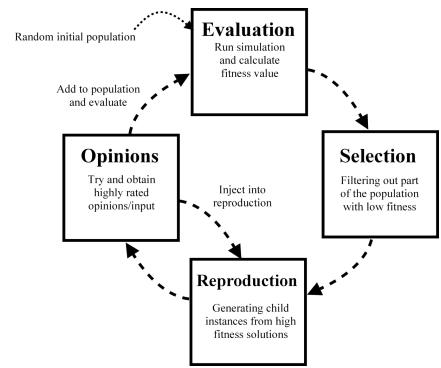

Figure 1: Optimization - Evolutionary optimization steps.

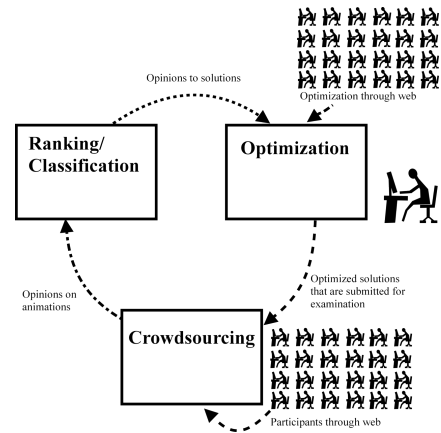

Figure 2: Cycle - Local optimization and the broader opinion based reviewing of the animation quality.

**Our Work** This paper is the first to bring together online crowdsourcing and web-based methods to tune and refine automatically generated character animations. The approach combines machine learning techniques (neural networks) with optimization approaches to create control signals for physically-based articulated characters. The animations are optimized using fitness criteria for core actions (walking, standing and jumping), however, the crowdsourcing element enables a vast number of online users to contribute to the aesthetic (organic and natural) component of the animation through a rating system.

## 3 Method

**Explore** Manipulating and coordinating of an articulated figure's limbs is a complex task which is accomplished via control signals [31]. We wanted the approach to be able to present ideas outside of the animators or designers imagination. To stimulate creative thought while presenting a usable solution. Having animation solutions adapt to new actions and present ideas to deal with unforeseen circumstances was important. A client-based solution that was not dependent on an internet connection to allow users to test out animation designs based on fitness criteria (control constraints).

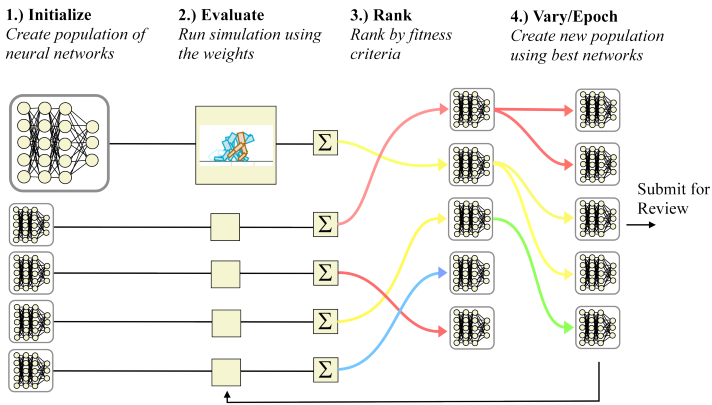

Figure 3: Simulation Steps - Population of neural networks are evaluated and evolved.

**Component-Based/Modular** Our implementation is written in a modular manner so that components, such as, the optimization algorithm, fitness function or skeleton topology could easily be swapped out or modified. Since the implementation is written in Javascript and runs locally in a browser, should be straightforward to expand and improve the implementation (given basic scripting skills). Flexible components/elements of the implementation included:

- Support and allow the expansion of multiple optimization algorithms (tested with genetic and differential algorithm for this implementation)
- Users might want to try different fitness criteria
- Different skeleton structures (bipeds, dinosaurs and hoppers)
- Noise generators (built in Javascript library vs custom versions that use Chaos concepts for better search/pattern identification)

Implementation steps

1. Javascript web-based solution (run standalone in a web-browser)
2. Physically-based simulation using Box2D
3. Neural network for the motion patterns
4. Evolutionary algorithm to train the neural network (fitness criteria for the mechanistic properties, such as motion, balance and goal)
5. Crowdsourcing to evaluate the generated animations (naturalness/realism/aesthetics).

Figure 1 and Figure 2 show a high level view of the optimization details (how feedback is fed into the tuning/refinement of motions through the solicitation of input from volunteers).

Key components/details of the implementation:

- Client-side (Javascript), server-side (Php)
- Neural network (Javascript SynapticJS library)
- Evolutionary search for refining the neural network (genetic or differential evolutionary algorithm)
- Simulation engine for collision/physics (Box2D)
- Implementation/optimization could run real-time on the client-side browser (however, only run local optimizations and would not inject subjective peer influences from other online crowdsourcing contributions)
- Store crowdsourcing ratings on animation quality online (server side)

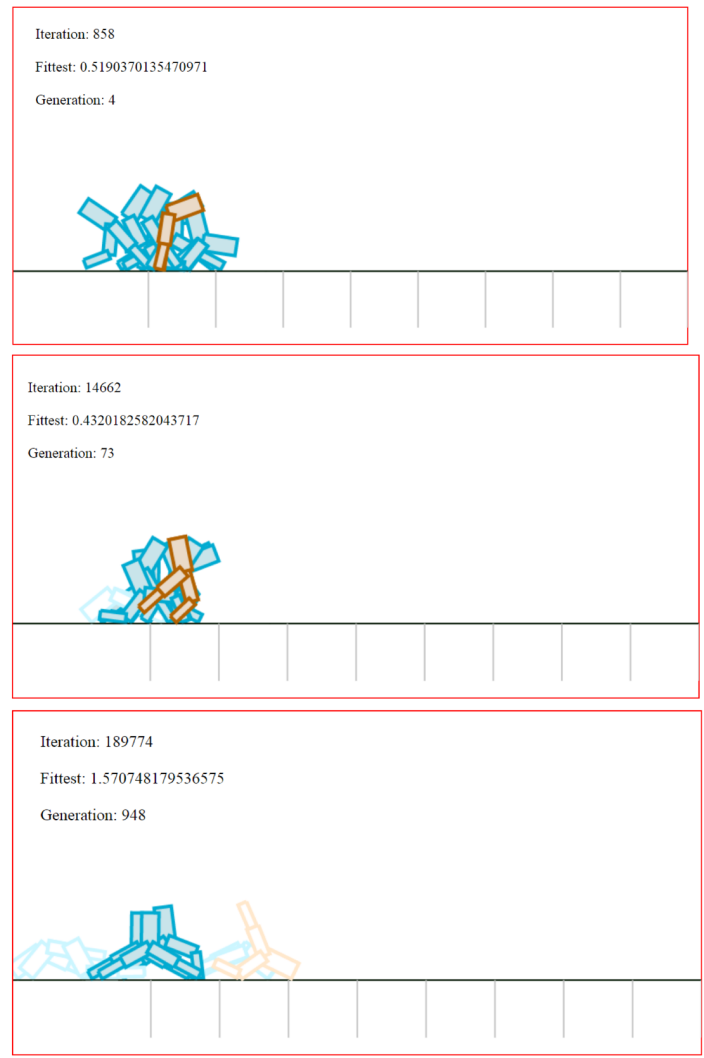

Figure 4: Walker Model Simulation - 5-limbs and 4-joints required to *walk* forward (towards the right). The optimization algorithm will create an assortment of solutions, jiggling, wiggling, hopping and flipping to move forward (from left-to-right). While these are *physically correct* they are not natural or the normal way of walking as accepted by society.

**Input & Output Variables** The input and output variables are crucial parameters for neural networks. The input variables consisted of the physical parameters for the articulated skeleton and the total time duration (since the start of the simulation). The physical parameters included the position (centre of mass), linear velocity and angular velocity of each rigid body. We included the timing information, as some motions might be oscillatory (not about getting the furthest but moving in a specific pattern). The output variables for the system was the control torques (fed into the physics simulator). The input parameters were normalized to a unit range (0-1) while the output parameters were scaled to an appropriate torque control limits.

The basic test walker neural network had 3 hidden layers. 21 inputs, 3 sets of 30 hidden nodes and 4 outputs (21:30:30:30:4).

**Why an Evolutionary Algorithm?** Multilayered feed-forward neural networks coupled with an evolutionary algorithm can really accelerate the learning process to solve certain problems [32] (known as evolutionary artificial neural networks - EANN). Evolutionary algorithms, such as the genetic algorithm, are bio-inspired optimization approaches that utilize populations of entities (best solution out of a pool of solutions - global optimal for a problem no known solution). Through evolutionary algorithms neural networks are able to solve a huge challenge (i.e., finding the necessary 'hyperparameters' for the network). Note, hyper-parameters are values required by the neural network to perform properly, given a problem. Neural networks possess a number of properties which make them particularly suited for generating complex animation patterns.However, their application to some real-world problems has been hampered by training algorithm which reliably finds a nearly globally optimal set of weights in a relatively short time. Evolutionary algorithms are a class of optimization procedures which are good at exploring a large and complex space in an intelligent way to find values close to the global optimum. Hence, they are well suited to the problem of training feed-forward networks. For details on how evolutionary algorithms operates for tuning animation signals (such as the genetic algorithm and differential evolutionary algorithm) see Kenwright [19, 20] for further details and explanations.

**Database & Crowdsourcing** Animation network solutions/ratings are stored online, with solutions 'deemed'. While users can run optimizations and rate their solutions, the database is also open, allowing external participants to view and rate individual solutions (without performing any optimizations). The judgement regarding whether an animation solution in the database was well or poorly defined by the 'wisdom of the crowd', crowdsourcing the opinions of external participants. Crowdsourcing, through the presentation of a 3D simulation. The overall assigned value for each solution in the database was the average rating.

The simulation developed for this purpose was developed using open source libraries that would run easily in a browser (Javascript, Box2D, 'Canvas'), and was available online (web-page), so as to facilitate easy accessibility for external participants. Participants could view previous simulations/views and optionally move on to run optimizations (simulations) of their own to create/add to the library.

Users were not offered judging standards but were instead asked if they found the best animation solution 'natural and life-like'. This approach was taken in an effort to **avoid the biased opinions** which would have probably resulted from excessive instructions. The assumption was that most of the animations would not receive a unanimous vote on their quality; consequently, the votes, about ten per unit, were eventually aggregated into a single average value, ranging between 0.0 and 1.0, which represented the crowdsourced opinion of the configuration of each unit. For machine learning purposes in this specific paper, the class of each unit was a binary value, Zero or One; animations with an averaged class larger than a chosen factor value, 0.5, were classified as good quality, the rest as poor. The value 0.5 represents an effective value (greater than or equal to half).

**Software/Libraries** The implementation was implemented in Javascript and PHP (sever) using the following libraries:

- Neural network (SynapticJS library [33]) About: Synaptic is a javascript neural network library for Node.js and the browser, its generalized algorithm is architecture-free, so you can build and train basically any type of first order or even second order neural network architectures
- Box2d Library [34]
- Web-based physics-engine written in Javascript
- Visual display/output (HTML5/Canvas)
- Uploading server side management was done with PHP

## 4 Experimental Results

A simple 'walking' model was the base-line test configuration and had an uncomplicated fitness criteria that was based-upon distance to the right (see Figure 4 (also the height of the pelvis must not drop below a threshold). Early generations would jump or fling themselves forward, however, after a large number of generations, the optimization search moves towards **rhythmic patterns** creating oscillatory joints signals propelling the character further (simple locomotion). This is where simple **fitness** criteria starts to fail, as it is possible to create motions that move very quickly but to not appear smooth and natural.

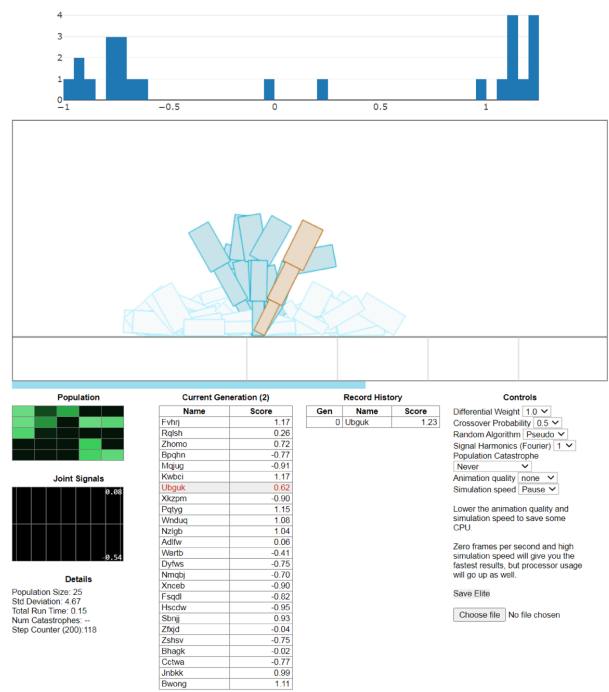

**Figure 5: Expanded options and visual statistics - Web-based implementation - with added features for the user/tester: Select different skeleton topology; Population distribution/fitness; Save/load current configuration; Optimization algorithm; Noise/random number algorithm;**

Additional criteria can be added to the fitness function, such as, maximum foot height and posture orientation, however, the more complex the fitness function the more it is bespoke to the specific action (e.g., walking vs running or climbing) - since, different fitness functions would need to be defined for the action, skeleton and the circumstances. This can be avoided by letting users 'rate' simple motions to assess their aesthetic qualities (help refine the search). As the simulation in its default form is limited in the amount of information given to the user (e.g., population diversity and fitness ratings), it is straightforward add addition user feedback details to the web-based interface (see Figure 5).

**Limitations** The current system has a number of limitations but demonstrates a proof of concept (for basic motions/actions). Our approach faces the same **human-centric challenges** of other open online systems, such as, how to recruit and evaluate users and combine their contributions [28]. The tests where only carried out on 2D character models, with the optimizations taking place locally in the browser (each optimization takes place on a single machine) and could have performance bottlenecks for much more complex skeletal structures/simulations. However, no optimizations or leveraging of GPU speedups was used, which have been shown to accelerate optimization algorithms [19].

## 5 General Discussion and Conclusion

The potential opportunity is presented in this paper is enormous, the ability to **tap into hundreds of thousands or millions of helpers** across the globe to accomplish creative work on an unprecedented scale. We have demonstrated a real-world application of an approach to a large and complex problem. We have also shown how adding option specific feedback into the loop can enhance and address uncanny animation attributes. Our system has the added advantage of being able to explore animation actions (physical and behavioural elements with the underpinning emotional aspects). Previous research into the field of animation optimization have emphasized the difficulties of implementing fitness criteria that capture organic life-like solutions (without pre-recorded animation data) [19, 35]. The work described here only touches the surface of the potential for using crowdsourcing to develop autonomous self-driven animation solutions.

**Future Work** The crowdsourcing approach used to rate animations in this paper only taps into the lowest levels of human intelligence. Future opportunities could be investigated to extend the crowdsourcing approach for complex tasks that require significant intelligence or creativity. Possibly using incentives and rewards to encourage user participation, even integrating the crowdsourcing work into a reward based-structure like a video game (gamification [36]).

Additional work could be done to create a neural network to perform **automatic classification** (i.e. supervised learning). Train a network model to distinguish between poor and high quality animations, on the basis of human evaluations. This future model would be based on supervised machine learning that includes periodic process of optimization, database expansion, crowdsourcing and classification. Finally, as a general-purpose solution the approach should be applicable to any type of neural network (and not just feedforward networks as used in this paper). Also possible to use training algorithms to aid in the development of other types of neural networks for the system.